\documentclass[preprint,aps]{revtex4}
\usepackage{amsmath}
\usepackage{amssymb} 

\begin{document} 
\bibliographystyle{apsrev} 
 
\title{Analysis of Grover's quantum search algorithm \\
as a dynamical system} 
\date{\today} 
 
\author{Ofer Biham, Daniel Shapira and Yishai Shimoni} 
\affiliation{Racah Institute of 
Physics, The Hebrew University, Jerusalem 91904, Israel} 
 
\begin{abstract} 
Grover's quantum search algorithm
is analyzed
for the case in which
the initial state 
is an arbitrary pure quantum state 
$| \phi \rangle$ 
of $n$ qubits. 
It is shown that the optimal time to perform the measurement
is independent of 
$| \phi \rangle$,
namely, it is identical to the optimal time in the original algorithm
in which
$| \phi \rangle = | 0 \rangle$,
with the same number of marked states, $r$.
The probability of success $P_{\rm s}$ 
is obtained,
in terms of the amplitudes of the state
$| \phi \rangle$,
and is shown to be independent of $r$.   
A class of states, which includes fixed points and cycles of the Grover
iteration operator is identified.
The relevance of
these results in the context of using the success probability as an
entanglement measure is discussed.
In particular, the Groverian entanglement measure,
previously limited to a single marked state,
is generalized to the case of several marked states.
\end{abstract} 
 
 
\maketitle 
 
\section{Introduction} 

Grover's quantum search algorithm 
\cite{Grover1996,Grover1997a} 
exemplifies
the potential speed-up offered by quantum computers. 
It also provides a
laboratory for the analysis of quantum algorithms and their 
implementation. 
The problem addressed by Grover's algorithm can be viewed as trying to 
find a marked element in an unsorted database of size $N$. 
While a classical computer would need, on average, 
$N/2$ database queries (and $N$ queries in the worst case)  
to solve this problem,
a quantum computer 
using Grover's algorithm, would 
accomplish the same task using merely 
$O(\sqrt{N})$ queries. 
The importance of Grover's 
result stems from the fact that it proves the enhanced power of
quantum computers compared to classical ones for a whole class of
oracle-based problems, for which the bound 
on the efficiency of classical
algorithms is known. 
Moreover, it was shown 
\cite{Zalka1999} 
that Grover's algorithm is as 
efficient as theoretically possible 
\cite{Bennett1997}.   
A variety of applications were developed, in which the
algorithm is used in the solution of other
problems
\cite{Grover1997b,Grover1997c,Terhal1998,Brassard1998,Cerf2000,Grover2000,Carlini1999}.
Experimental implementations 
of Grover's algorithm
were constructed using 
nuclear magnetic resonance 
(NMR)
\cite{Chuang1998b,Jones1998}
as well as an optical device
\cite{Kwiat2000}. 

Several generalizations of Grover's original
algorithm have been developed. 
The case in which there are several marked 
states was studied in 
Ref.~\cite{Boyer1996}. 
It was shown that when there are
$r$ marked states, 
Grover's algorithm would be able to find one of them
after 
$T=O(\sqrt{N/r})$ queries.
A further generalization was obtained 
by allowing 
the replacement of the Hadamard transform, used in the original setting,
by an arbitrary (but constant) unitary
transformation 
\cite{Grover1998a,Gingrich2000,Biham2001},
as well as by the replacement of the $\pi$ inversion 
by an arbitrary (but constant) phase rotation 
\cite{Long1999a}. 

Another generalization was obtained 
by allowing the replacement of the 
uniform superposition of all basis states,
used as the initial state of the algorithm
in the original setting, 
by an arbitrary pure 
\cite{Biron1997,Biham1999} 
or mixed
\cite{Biham2002e}
quantum state.
In this case the probability of success 
of the algorithm, 
$P_{\rm s}$, 
may be
reduced, depending on 
the initial state.
Recently it was shown that this generalization gives rise 
to an entanglement monotone that can be used to quantify the
entanglement in pure states of multiple qubits
\cite{Biham2002}.

In this paper we provide a comprehensive analysis of the
dynamical behavior and the 
success probability of the quantum search algorithm 
for arbitrary initial pure quantum states.
We show that for a given initial state, 
$| \phi \rangle$,
the success probability does not depend on the number
of marked states used in the algorithm.
We provide an explicit expression for $P_s$
in terms of the amplitudes of the initial state, and
discuss a simple geometrical interpretation of it.
We then use this approach to calculate $P_{\rm s}$ for states
that exhibit various symmetries, as well as states that
typically appear in quantum algorithms. 
Two special classes of states are identified:
fixed points, namely states that remain invariant under the
Grover iteration as well as two-cycles.
The implications of the results in the context of 
entanglement measures for quantum states of multiple qubits are discussed.

The paper is organized as follows. 
In Sec. II we briefly describe the algorithm. 
The algorithm in the case of
an arbitrary initial quantum state is studied in Sec.
III, where an expression for $P_s$ is obtained
and shown to be independent of $r$,
and fixed points and cycles are identified.
The results are discussed in Sec. IV and
summarized in Sec. V.
 
\section{The quantum search algorithm} 
\label{sec:algorithm} 
 
Consider a search space $D$ containing $N$ elements.  We assume, for 
convenience, that $N = 2^n$, where $n$ is an integer. In this way, we 
may represent the elements of $D$ using an $n$-qubit {\em register} 
containing the indices, $i=0,\dots,N-1$.  We assume that a subset of 
$r$ elements in the search space are marked, that is, they are 
solutions of the search problem.  The distinction between the marked 
and unmarked elements can be expressed by a suitable function, 
$f: D \rightarrow \{0,1\}$, 
such that $f=1$ for the marked elements, and $f=0$ for the rest. 
The search for a marked element now becomes a search for an element 
for which $f=1$.  To solve this 
problem on a classical computer one needs to evaluate $f$ for each 
element, one by one, until a marked state is found.  Thus, on average, 
$N/2$ evaluations of $f$ are required and $N$ in the worst case.
For a quantum computer, on which
$f$ to be evaluated 
\emph{coherently}, 
it was shown that a sequence of unitary operations 
called Grover's algorithm
can locate a marked element using only $O(\sqrt{N/r})$ coherent 
queries of $f$ \cite{Grover1996,Grover1997a}.  
 
To describe the operation of the quantum search algorithm we first 
introduce a register, 
$\left| i \right\rangle = 
\left| i_{1} \ldots   i_{n} \right\rangle$, 
of $n$ qubits, 
and an 
\emph{ancilla} qubit, 
$|q\rangle$, 
to be used in the computation.  
We also introduce a 
\emph{quantum oracle}, 
a unitary operator $O$ which functions as a 
black box with the ability to 
\emph{recognize} solutions to the search 
problem.  
The oracle performs the following 
unitary operation on computational basis states of the register, 
$\left| i \right\rangle$, 
and the ancilla, 
$\left| q \right\rangle$:
 
\begin{equation}
O \left| i \right\rangle \left| q \right\rangle = 
\left| i \right\rangle   \left| q \oplus f(i) \right\rangle 
\label{eq:bborac} 
\end{equation} 

\noindent
where $\oplus$ denotes addition modulo 2. 
The oracle recognizes marked states in the sense that if 
$| i \rangle$ 
is a marked element of the search space, 
namely $f(i) = 1$, 
the oracle flips the ancilla qubit from 
$\left| 0 \right\rangle$ 
to 
$\left| 1 \right\rangle$ 
and vice versa, 
while for unmarked states the ancilla is unchanged.  
The ancilla qubit is initially 
set to the state 

\begin{equation} 
| - \rangle_q = { \frac{1}{\sqrt{2}} } (\left| 0 \right> - \left| 1 \right>).  
\end{equation} 

\noindent
With this choice, the action of the oracle is: 

\begin{equation} 
O |i\rangle |-\rangle 
= (-1)^{f(i)} |i\rangle |-\rangle_q.
\end{equation} 

\noindent
Thus, the only effect of the oracle is to apply a phase of $-1$ if $x$ 
is a marked basis state, and no phase change if $x$ is unmarked.  
Since the state of the ancilla does not change, 
one my omit it and write the action of the oracle as 
$O|x\rangle = (-1)^{f(x)}|x\rangle$.  

Grover's search algorithm may be described as follows: 
Given a black box oracle $O$, whose action 
is defined by Eq.~(\ref{eq:bborac})
and a register of $n+1$ qubits in the 
state $|0\rangle^{\otimes n}|0\rangle_q$,
the following procedure is performed:

\begin{enumerate} 
\item \label{en:initst} Initialization: 
Apply a Hadamard gate    
$H = \frac{1}{\sqrt2}
\left(
\begin{smallmatrix} 
1 & 1 \\ 
1 & -1 
\end{smallmatrix}
\right)$ 
to each qubit in the register, 
and the gate $HX$ to the ancilla, 
where 
$X=
\left(
\begin{smallmatrix} 
0 & 1 \\ 
1 & 0 
\end{smallmatrix}
\right)$ 
is the {\sc not} gate,  
and we write matrices with respect to the computational basis 
($|0\rangle,|1\rangle$).  
The resulting state is: 

\begin{equation}
|\eta \rangle |-\rangle_q, 
\label{eq:etam} 
\end{equation} 

\noindent
where

\begin{equation}
| \eta \rangle = \frac{1}{\sqrt{N}} \sum_{i=0}^{N-1} |i\rangle. 
\label{eq:eta} 
\end{equation} 

\noindent
\label{init:item} 
\item Grover Iterations: Repeat the following operation $\tau$ times 
  (where $\tau$ is an integer given below). 
\begin{enumerate} 
\item \label{en:rot1} Apply the oracle, which has the effect of 
  rotating the marked states by a phase of $\pi$ radians.  Since the 
  ancilla is always in the state $|-\rangle_q$ the effect 
  of this operation 
  may be described by the following unitary operator 

\begin{equation}
I_f^{\pi} = \sum_{i=0}^{N-1} (-1)^{f(i)} | {i} \rangle \langle {i} |, 
\end{equation} 

acting only on the register. 

\item (i) apply the 
  Hadamard gate on each qubit in the register; (ii) Rotate the 
  $\left| 00 \ldots 0 \right\rangle$ state  
  of the register by a phase of $\pi$ 
  radians.  This rotation is similar to 2(a), except for the fact that 
  here it is performed on a known state. It takes the form
 
\begin{equation}
I_{0}^{\pi} = - 2 |0\rangle \langle 0| + \sum_{i=0}^{N-1} |i\rangle \langle i|, 
\end{equation} 

where the second term on the right hand side is the identity operator, denoted
by $I$; 
(iii) Apply the Hadamard gate again on each qubit in the register. 

The resulting operation is

\begin{equation}
- H^{\otimes n} I_{0}^{\pi} H^{\otimes n} = 
    -I + 2 H^{\otimes n} |0\rangle \langle 0| H^{\otimes n} =
    -I + | \eta \rangle \langle \eta |.
\end{equation}

When this operator is applied on the state 
$\sum_i a_i | i \rangle$
it results in the state 
$\sum_i (2 \bar{a} - a_i) | i \rangle$,
where $\bar a = \sum_i a_i/N$. 
Thus, each amplitude is rotated by $\pi$ around the
average of all amplitides of the quantum state.
\end{enumerate} 

The combined operation on the register
in one Grover iteration 
is described by 

\begin{equation}
U_G = - H^{\otimes n} I_{0}^{\pi} H^{\otimes n} I_f^\pi. 
\label{eq:U_G}
\end{equation}

\item  Measure the register in the computational basis. 
\end{enumerate} 

\noindent
The optimal number of iterations is

\begin{equation}
\tau = \left\lfloor
\frac{\pi}{4} \sqrt{\frac{N}{r}}
\right\rfloor, 
\label{eq:optit} 
\end{equation} 

\noindent
where $\lfloor x \rfloor$
is the largest integer which is smaller than $x$
\cite{Grover1997a,Boyer1998,Zalka1999}. 
Moreover, at this optimal time a marked state
can be found with almost certainty, or more 
precisely with probability

\begin{equation}
P_{\rm s} = 1 - O \left({ \frac{1}{\sqrt{N}} } \right). 
\end{equation}

\noindent
With this performance,
Grover's algorithm was found to be 
optimal in the sense that it is as efficient as 
theoretically possible
\cite{Bennett1997}. 

Note that the probability 
$P \approx 1$ 
can be achieved only 
for specific initial states such as the one 
produced in step~$1$ of the algorithm above.  
If this starting state is replaced by 
an arbitrary quantum state, the
probability of succes, $P_{\rm s}$, is reduced
\cite{Biron1997,Biham1999}. 
In the next Section we 
analyze the operation of the algorithm in the case
that the initial state is an arbitrary pure
quantum state
$| \phi \rangle$.
The time evolution of the quantum state
during the Grover iterations is examined,
for a specific choice of the marked states.
Special states that exhibit fixed points and cycles
under Grover iterations are identified.  
A closed form expression for 
$P_{\rm s}$ is obtained,
and a geometrical interpretation for it is presented.
It is shown that for a given state
$| \phi \rangle$, the success probability does not
depend on the number of marked states.
Finally, some specific quantum states are examined
and their utility as initial states for Grover's algorithm
is obtained.

\section{Grover's algorithm with an arbitrary pure initial state}

\subsection{The initial state}

Consider an arbitrary pure quantum
state
$| \phi \rangle$
of $n$ qubits,
to be used as the initial state in Grover's algorithm.
It can be expressed by

\begin{equation}
| \phi \rangle = \sum_{i=0}^{N-1} a_i(0) | i \rangle,
\label{eq:phi}
\end{equation}

\noindent
where the amplitudes
$a_i(0)$, $i=0,\dots,N-1$
are complex numbers that satisfy

\begin{equation}
\sum_{i=0}^{N-1} |a_i(0)|^2 = 1,
\label{eq:normalization}
\end{equation}

\noindent
and $N=2^n$.
The distribution of these amplitudes can be characterized by
its moments.
The first moment (or average) of the amplitudes
is

\begin{equation}
\overline{a(0)} = {\frac{1}{N}} \sum_{i=0}^{N-1} a_i(0). 
\label{eq:a_bar}
\end{equation}

\noindent
The second moment  

\begin{equation}
\overline{|a(0)|^2} = {\frac{1}{N}} \sum_{i=0}^{N-1} |a_i(0)|^2
\end{equation}

\noindent
satisfies
$\overline{|a(0)|^2} = 1/N$,
for any state
$| \phi \rangle$,
due to the normalization condition
(\ref{eq:normalization}).
The standard deviation of the amplitude distribution 
is given by

\begin{equation}
\sigma_{a}^{2}(0) = \frac{1}{N} \sum_{i=0}^{N-1} |a_{i}(0) - \overline{a(0)}|^{2}.
\end{equation}

\noindent
Using the equality

\begin{equation}
\sigma_{a}^{2}(0) = \overline{|a(0)|^2} - \left|\overline{a(0)}\right|^2
\end{equation}

\noindent
we obtain that

\begin{equation}
\sigma_{a}^{2}(0) = {\frac{1}{N}} - \left|\overline{a(0)}\right|^2.
\label{eq:SigmaAv}
\end{equation}

\noindent
This result can be used to identify two limits.
One limit is the equal superposition state,
in which 
$a_i(0) = 1/\sqrt{N}$
for
$i=0,\dots,N-1$.
In this state
$\overline{a(0)} = 1/\sqrt{N}$
and 
$\sigma_a(0) = 0$.
In the opposite limit there is a large family
of states for which
$\overline{a(0)} = 0$.
In this case
$\sigma_a(0) = 1/\sqrt{N}$,
which is
the largest value that  
$\sigma_a(0)$
can take.

\subsection{Recursion equations for a pre-defined set of marked states}

Consider Grover's algorithm,
searching for one of $r$ marked states
where the
initial state is
$| \phi \rangle$.
Denote the set of indices of the marked states
by ${\rm M}$.
The amplitudes of the marked states
will thus be
$a_i$, $i \in {\rm M}$.
In some cases below, we will denote the marked 
states by
$|m_1\rangle,\dots,|m_r\rangle$
and their amplitudes by
$a_{m_1}(0),\dots,a_{m_r}(0)$.
The complementary set, which includes the indices 
of the unmarked states, will be denoted by
$\overline{\rm M}$.

The time evolution of the amplitudes of the marked and
unmarked states during the Grover's iterations 
with an arbitrary pure quantum state as the initial state,
was studied in Ref.
\cite{Biham1999}.
Starting from the state 
$| \phi \rangle$,
the amplitudes obtained after $t$ Grover iterations 
are denoted by $a_i(t)$, $i=0,\dots,N-1$.

For a given choice of the marked states, one can
consider separately the averages and standard deviations
of the sets of marked and unmarked states at time $t$.
The averages will be

\begin{equation}
\overline{a_m(t)} = {\frac{1}{r}} \sum_{i \in M} a_{i}(t) 
\label{eq:am_bar}
\end{equation}

\noindent
for the marked states,
and

\begin{equation}
\overline{a_u(t)} = {\frac{1}{{N-r}}} \sum_{i \in \overline{\rm M}} a_i(t) 
\label{eq:aum_bar}
\end{equation}

\noindent
for the unmarked states.
The standard deviations will be

\begin{equation}
\sigma_{m}^{2}(t) = \frac{1}{r} \sum_{i \in {\rm M}} |a_{i}(t) - \overline{a_m(t)}|^{2}
\end{equation}

\noindent
for the marked states,
and

\begin{equation}
\sigma_{u}^{2}(t) = \frac{1}{N-r} 
\sum_{i \in \overline{\rm M}} |a_{i}(t) - \overline{a_{u}(t)}|^{2}.
\end{equation}

\noindent
for the unmarked states.

Each Grover iteration consists of two steps. In the
first step the phases of all the marked amplitudes are
totated by $\pi$, 
namely
$a_{i} \rightarrow - a_{i}$, $i \in {\rm M}$.
In the second step all the amplitudes are
rotated by $\pi$
around their average,
namely
$a_i \rightarrow 2 \bar{a} - a_i$, $i=0,\dots,N-1$.
Using these properties,
the time dependence was found to be described 
by the recursion equations
\cite{Biham1999}

\begin{eqnarray}
a_{i}(t+1) &=& C(t) +  a_{i}(t)  \ \ \ i \in {\rm M} \nonumber \\
a_i(t+1)     &=& C(t) -  a_i(t)  \ \ \ i \in \overline{\rm M}
\label{eq:ampvst}
\end{eqnarray}

\noindent
and

\begin{equation}
C(t) = {\frac{2}{N}} \left[ (N-r) \overline{a_u(t)} - r \overline{a_m(t)} \right].
\label{eq:C(t)}
\end{equation}

\noindent
It was also shown that the standard deviations
of the amplitude distributions of the marked
and unmarked states are constants of motion,
namely
$\sigma_m(t) = \sigma_m(0)$
and
$\sigma_u(t) = \sigma_u(0)$
at any time $t$.

\subsection{Solution of the recursion equations}

Consider a quantum search with $r$ marked states,
using an arbitrary quantum state 
$| \phi \rangle$
as the initial state.
It was shown in Ref.
\cite{Biham1999}
that the time evolution of the amplitudes is given by

\begin{eqnarray}
a_i(t) &=& \overline{a_m(t)}+\Delta a_i:  \ \ \ \ \ \ \ \ \    i \in {\rm M} \nonumber \\
a_i(t) &=& \overline{a_u(t)}+ (-1)^t \Delta a_i: \ \ \ i \in \overline{\rm {M}},
\label{eq:aioft}
\end{eqnarray}

\noindent
where

\begin{eqnarray}
\overline{a_m(t)} &=&  \sqrt{\frac{N-r}{r}} \alpha \sin(\omega t + \delta) \nonumber \\
\overline{a_u(t)} &=&  \alpha \cos(\omega t + \delta)
\label{eq:abaroft}
\end{eqnarray}

\noindent
and

\begin{eqnarray}
\Delta a_i &=& a_i(0) - \overline{a_m(0)}: \ \ \ \ \ \    i \in {\rm M} \nonumber \\
\Delta a_i &=& a_i(0) - \overline{a_u(0)}: \ \ \ \ \ \ \    i \in \overline{\rm M}.
\label{eq:daioft}
\end{eqnarray}

\noindent
The parameters $\alpha$
and
$\delta$
are given by

\begin{eqnarray}
\alpha &=& \sqrt{\overline{a_u(0)}^2 + {\frac{r}{{N-r}}} \ \overline{a_m(0)}^2 } \nonumber \\
\exp{(2 i \delta)} &=& 
{ {\sqrt{N-r} \  \overline{a_u(0)} + i \sqrt{r} \ \overline{a_m(0)}}
\over
{\sqrt{N-r} \  \overline{a_u(0)} - i \sqrt{r} \ \overline{a_m(0)}} },
\label{eq:phieq}
\end{eqnarray}

\noindent
where
$-\pi/2 \le {\rm Re}(\delta) < \pi/2$.
Furthermore, it was found 
that if a measurement is taken after
$t$ iterations, the probability to find one
of the marked states is given
by

\begin{equation}
P(t) = P_0 - \Delta P \cos^2[\omega t + {\rm Re} (\delta)]
\label{eq:Pt}
\end{equation}

\noindent
where

\begin{equation}
P_0 = 1 - (N-r) \sigma_u^2 - {\frac{1}{2}} K
\label{eq:P_0}
\end{equation}

\noindent
is the highest probability of success that can be achieved for the
specific choice 
of the set of marked states
denoted by
${\rm M}$ 
(if the measurement is taken at the optimal time).
The parameter $K$ takes the form

\begin{equation}
K = (N-r) \left|\overline{a_u(0)}\right|^2 
  + r \left|\overline{a_m(0)}\right|^2 
  - \left|(N-r) {\overline{a_u(0)}}^2 + r {\overline{a_m(0)}}^2  \right|. 
\label{eq:CinP}
\end{equation}

\noindent
The coefficient of the time dependent term is

\begin{equation}
\Delta P =  \left|(N-r) {\overline{a_u(0)}}^2 
         + r {\overline{a_m(0)}}^2  \right|. 
\end{equation}

\noindent
The frequency $\omega$ given by

\begin{equation}
\cos \omega = 1 - {\frac{2r}{N}},
\end{equation}

\noindent
or approximately by
$\omega = 2 \sqrt{r/N}$
where $r \ll N$.
The function $P(t)$ is a sinusoidal function. In the analysis of the efficiency
of the algorithm we are interested in the largest value 
$P_0$
that this function may
reach during its cycle. 
We are also interested in the number of iterations that it would take
to reach this maximum for the first time, from the given initial state.
This optimal number of iterations,
for the given set, ${\rm M}$ of marked states
is given by

\begin{equation}
\tau_{\rm M} = 
\left\lfloor {\frac{1}{2}} \sqrt{\frac{N}{r}} 
\left( {\frac{\pi}{2}} - {\rm Re}(\delta)  \right) \right\rfloor.
\label{eq:t_opt}
\end{equation}

\noindent
Note that for 
$0 \le {\rm Re}(\delta) < \pi/2$,
$\tau_{\rm M} \le \tau$
namely the optimal number of iterations is equal or smaller than for
the initial state
$| \eta \rangle$.
On the other hand,
for 
$-\pi/2 \le {\rm Re}(\delta) < 0$
the optimal number of iterations
$\tau_{\rm M}$
is larger than
$\tau$.
Note that within this analysis, in which the set of marked states is assumed to be known,
it would be more efficient 
for
$-\pi/2 \le {\rm Re}(\delta) < 0$,
to apply the Grover ierations
(\ref{eq:U_G})
in reverse order.
Of course this does not apply to the actual search in which the marked
states are unknown.

There are several issues to consider about the probability
$P_0$.
First, the case that we considered so far is
not the real search. This is due to the
fact that in the derivation of 
Eq.~(\ref{eq:Pt})
it was assumed that the identities of the marked states
and their amplitudes are known.
In order to obtain the probability of success of an 
actual search one needs to average
Eq.~(\ref{eq:Pt})
over all possible
sets of the $r$ marked states.
In the case of one marked state this yields an average
over $N$ possibilities.
In general, for $r$ marked states the number of possible sets
is given by the binomial coefficient
$\left(
\begin{smallmatrix} 
N  \\ 
r  
\end{smallmatrix}
\right)$.
Averaging
over all these 
possibilities
yields
$\langle \overline{a_{\rm m}(0)} \rangle$,
which is the average of
$\overline{a_{\rm m}(0)}$
over all possible sets 
of $r$ marked states.
This average is used as a small parameter in
the series expansion that provides the
average of $P(t)$ over all these possible choices of the marked
states.

Consider the constant $K$
in Eq.
(\ref{eq:CinP}).
For an initial state 
$| \phi \rangle$
in which all the amplitudes are 
real, there is an exact cancellation
and $K=0$.
Clearly, this simplifies the calculation of
$P_0$ significantly.
In case that the amplitudes are complex,
and the phases of
$\overline{a_m(0)}$
and 
$\overline{a_u(0)}$
are different,
$K$ may be nonzero.
It is easy to see that it is bounded by
$0 \le K \le \min \{ 2 r \left|\overline{a_m(0)} \right|^2, 
2 (N-r) \left| \overline{a_u(0)} \right|^2 \}$.
Consider, for example, the case of a single marked state.
For a specific choice of $m_1$, the constant $K$ can take
any value in the range $0 \le K \le 1$.
However, averaging it over all the $N$ possible choices
of the marked state, it is easy to see
that $K$ is reduced to order $1/N$.

For a given choice of the set of marked states,
the optimal time to measure is found to depend on
the initial state according to
Eq.
(\ref{eq:t_opt}).
However, by averaging $P(t)$ 
[Eq. (\ref{eq:Pt})]
over all possible choices of
the set of marked states, 
Eq.
(\ref{eq:P_0})
is replaced
by

\begin{equation}
P_0 = 1 - N \sigma_a^2 + O\left(\frac{1}{\sqrt{N}}\right), 
\label{eq:P_0aver}
\end{equation}

\noindent
and using Eq.
(\ref{eq:SigmaAv})
we obtain

\begin{equation}
P(t) = N  \left| \overline{a(0)} \right|^2 \sin^2 (\omega t) 
+ O\left(\frac{1}{\sqrt{N}}\right). 
\label{eq:Ptaver}
\end{equation}

\noindent
Therefore, for any
initial state 
$| \phi \rangle$, 
the optimal number of iterations to be performed before 
taking the measurement remains the same, namely $\tau$
[Eq. (\ref{eq:optit})].
The probability of success is 
$P_{\rm s} = N \left| \overline{a(0)} \right|^2$.
The averaging process that leads to this result is rather tedius. 
In the next Sections we will present a more elegant
approach for the calculation of 
$P_{\rm s}$.

\subsection{Fixed points and cycles}

Typically, the function $P(t)$ is a quasi-periodic function since,
generically,
the frequency $\omega/(2 \pi)$ is incommensurate with the sequence 
of integers counting the Grover iterations.
A periodic (rather than quasi-periodic) cycle of the amplitudes 
is obtained when the frequency $\omega$ is a rational product
of $2 \pi$.
This condition is satisfied when

\begin{equation}
\cos^{-1}\left(1-2 {\frac{r}{N}}\right) = {\frac{p}{q}} \pi.
\end{equation}

\noindent
One example of such periodic cycle appears when $N/r=4$. In
this case $\omega=\pi/3$, and the cycle is of period 6.

For some initial states and specific choices of the set
of marked states, Grover's algorithm does not provide
any enhancement of the probability to obtain a marked
state. Such situations appear when
$\Delta P = 0$,
thus the success probability $P(t)$ becomes a constant.
The condition for 
$\Delta P = 0$
can be expressed by

\begin{equation}
\overline{a_m(0)} = \pm i \sqrt{ \frac{N-r}{r} } \overline{a_u(0)}. 
\label{eq:DeltaPis0}
\end{equation}

\noindent
Under this condition
the amplitude
$\alpha=0$,
and thus 
$\overline{a_m(t)}=0$
and
$\overline{a_u(t)}=0$.
It is easy to see from the geometrical description of the algorithm
that in this case
the operator $U_G$ exhibits cycles of period 2,
namely 

\begin{equation}
U_G^2 |\phi\rangle = |\phi\rangle.
\label{eq:twocycle}
\end{equation}

\noindent
The condition for a two-cycle is simply
$\overline{a_m(0)}=0$
and
$\overline{a_u(0)}=0$.
This is due to the fact that under this condition
the amplitudes of the marked states are invariant
under $U_G$,
while those of the unmarked states change sign at
each iteration
(what matters is the relative phase of $\pi$ between the
marked and unmarked states, while the global phase can be ignored).

We will also consider  
fixed points of $U_G$,
namely states that satisfy

\begin{equation}
U_G |\phi\rangle = |\phi\rangle.
\label{eq:fixedp}
\end{equation}

\noindent
The set of fixed points of
$U_G$ consists of two classes:
(a) states for which
$\overline{a_m(0)}=0$
and
$a_i(0)=0$ 
for each of the unmarked states;
(b) states for which 
${a_{m_i}(0)}=0$
for each of the marked states
and
$\overline{a_u(0)}=0$.
In the case of a single marked state, 
$| m_1 \rangle$,
a fixed point is obtained when
$a_{m_1}(0)=0$
and
$\overline{a_u(0)}=0$
(where at least two of the unmarked amplitudes are non-zero).

Consider the case of two marked states,
$| m_1 \rangle$
and 
$| m_2 \rangle$.
The state

\begin{equation}
| \phi \rangle = {\frac{1}{\sqrt{2}}} (|m_1\rangle - |m_2\rangle)
\label{eq:m1-m2}
\end{equation}

\noindent
is a fixed point of $U_G$
because the amplitudes of all the unmarked states vanish
and the average amplitude of the marked states vanishes as well.
In the case of $r$ marked states,
$| m_1 \rangle,\dots,| m_r \rangle$,
the states 

\begin{equation}
| \phi \rangle =  \sum_{i=1}^r a_{m_i}(0) |m_i\rangle,
\label{eq:m1..mr}
\end{equation}

\noindent
where
$\overline{a_m(0)} = 0$,
are fixed
points of $U_G$.

We conclude that with the exception of those periodic cycles with 
periods related to the frequency $\omega$, 
the only possible cycles of $U_G$ are the fixed points and cycles
of period 2, discussed above.

\subsection{The probability of success for one marked state}

The initialization step of the original algorithm
leads to the state
$| \eta \rangle$,
given by Eq.
(\ref{eq:eta}). 
Consider the case that there are $r$ marked states,
$| m_i \rangle$, $i=1,\dots,r$.
The operation of the algorithm, 
using the optimal initial state,
$| \eta \rangle$,
and performing the optimal number of iterations, $\tau$
can be expressed
by

\begin{equation}
U_G^{\tau} \left| \eta \right> = 
{\frac{1}{\sqrt{r}}} \sum_{i=1}^r | m_i  \rangle + 
O\left(\frac{1}{\sqrt{N}}\right). 
\label{eq:getms} 
\end{equation} 

\noindent
We will now examine the case in which the initial
state is 
an arbitrary pure quantum state
$| \phi \rangle$ of $n$ qubits.
The purpose of the algorithm is to find one of the
marked states, which are not known in advance.
As we have seen in the previous Section,
the optimal number of iterations is
$\tau$,
which is independent of the initial state
$| \phi \rangle$. 
The final state, just before the measurement of the
register, will thus be
$U_G^{\tau} \left| \phi \right> $.  
We will now calculate the probability that
a measurement of this state will yield
one of the marked states. 
For simplicity, we will first consider the case of 
one marked state
$| m_1 \rangle$.
In this case, for the initial state
$| \eta \rangle$
we obtain

\begin{equation}
U_G^{\tau} \left| \eta \right> = | m_1 \rangle + 
O\left(\frac{1}{\sqrt{N}}\right). 
\label{eq:Uetam1}
\end{equation} 

\noindent
For an arbitrary initial state
$| \phi \rangle$,
the resulting state will be
$U_G^{\tau} \left| \phi \right> $,
and the probability that a 
measurement of this state will 
yield the marked state
$| m_1 \rangle$
is

\begin{equation}
P_{\rm s} =  \left| \langle m_1 | U_G^{\tau} |\phi\rangle \right|^2  
+ O\left(\frac{1}{\sqrt{N}}\right). 
\label{eq:pmax2} 
\end{equation} 

\noindent
Note that Eq.
(\ref{eq:Uetam1})
can be written in the form

\begin{equation} 
\langle m_1| U_G^{\tau} = \langle\eta| + O\left(\frac{1}{\sqrt{N}}\right). 
\end{equation} 

\noindent
Inserting this equation into
(\ref{eq:pmax2})
we obtain that

\begin{equation} 
P_{\rm s} = \left| \langle    \eta | \phi     \rangle  \right|^2
+ O\left(\frac{1}{\sqrt{N}}\right),
\label{eq:pmaxetaphi}
\end{equation} 

\noindent
namely,  
$P_{\rm s}$
is given by the overlap of
$\phi$
with
the equal superposition $| \eta \rangle$.

Note that in the analysis above we used an arbitrary marked
state
$| m_1 \rangle$.
Since the marked state is not known, the calculation
of $P_{\rm s}$
should be done by averaging
Eq. 
(\ref{eq:pmax2})
over all the $N$ possible choices of the marked state.
It will thus take the form

\begin{equation}
P_{\rm s} =  \frac{1}{N} \sum_{i=0}^{N-1} 
\left| \langle i | U_G^{\tau} | \phi \rangle \right|^2  
+ O\left(\frac{1}{\sqrt{N}}\right), 
\label{eq:pmax3} 
\end{equation} 

\noindent
where in each of the $N$ terms the operator $U_G$ is 
constructed such that the marked state is the corresponding
state $| i \rangle$.
It turns out that this averaging is not really necessary,
because all the $N$ terms in this sum are identical
[up to corrections of order $O(1/\sqrt{N})$]. 

Eq.
(\ref{eq:pmaxetaphi})
can be used in order 
to express
the success probability 
$P_{\rm s}$
in terms of the amplitudes 
of
$| \phi \rangle$.
To this end we will 
express the states
$| \eta \rangle$
and
$| \phi \rangle$
in terms of their amplitudes,
according to Eqs.
(\ref{eq:eta}),
and 
(\ref{eq:phi}),
respectively,
and obtain

\begin{equation}
P_{\rm s} = N \left|\overline{a(0)}\right|^2 + O\left(\frac{1}{\sqrt{N}}\right).
\end{equation}
 
\noindent
Clearly, 
$P_{\rm s}$
turns out to depend only on the first moment
of the distribution of the amplitudes.
Initial states that exhibit high values of
$P_{\rm s}$
are those
for which the average amplitude
$\bar a$
is large (in absolute value).

\subsection{The probability of success for two marked states}

Consider the quantum search algorithm with an initial state
$| \phi \rangle$
and two marked states,
$| m_1 \rangle$
and 
$| m_2 \rangle$.
The probability 
that the algorithm will yield one of the marked
states is
   
\begin{equation}
P_{\rm s} =  
\left| \langle m_1 | U_G^{\tau} |\phi\rangle \right|^2  
+ \left| \langle m_2 | U_G^{\tau} |\phi\rangle \right|^2  
+ O\left(\frac{1}{\sqrt{N}}\right). 
\label{eq:pmax4} 
\end{equation} 

\noindent
In this case, 
the operation of the algorithm on the initial
state
$| \eta \rangle$
can be described by

\begin{equation}
U_G^{\tau} \left| \eta \right> = 
{\frac{1}{\sqrt{2}} } (|m_1 \rangle + |m_2 \rangle) 
+ O\left(\frac{1}{\sqrt{N}}\right). 
\label{eq:getms2} 
\end{equation} 

\noindent
or

\begin{equation}
{\frac{1}{\sqrt{2}} } (\langle m_1 | + \langle m_2 |) 
U_G^{\tau} 
= 
\langle \eta | 
+ O\left(\frac{1}{\sqrt{N}}\right). 
\label{eq:getmsl} 
\end{equation} 

\noindent
unlike the case of a single marked state, this equation
cannot be applied directly for the evaluation of $P_{\rm s}$.
The reason is that $P_{\rm s}$ is determined by the overlaps of
$U_G^{\tau} | \phi \rangle$ 
with each one of the marked states separately and 
not with their superposition.
To overcome this difficulty we will use the following identity:

\begin{equation}
\left| \langle m_1 | \psi \rangle \right|^2 +
\left| \langle m_1 | \psi \rangle \right|^2  
= 
{\frac{1}{2}} \left| (\langle m_1 | + \langle m_2 |) \psi \rangle \right|^2  +
{\frac{1}{2}} \left| (\langle m_1 | - \langle m_2 |) \psi \rangle \right|^2,  
\label{eq:identity}
\end{equation}

\noindent
where $| \psi \rangle$
is an arbitrary quantum state of $n$ qubits.
Using this identity we can write

\begin{equation}
\left| \langle m_1 | U_G^{\tau} |\phi\rangle \right|^2  
+ \left| \langle m_2 | U_G^{\tau} |\phi\rangle \right|^2  
=
{\frac{1}{2}} \left| (\langle m_1 | + \langle m_2 |) U_G^{\tau} 
|\phi\rangle \right|^2  
+ {\frac{1}{2}} \left| (\langle m_1 | - \langle m_2 |) U_G^{\tau} 
|\phi\rangle \right|^2.  
\label{eq:symasym}
\end{equation}

\noindent
In both terms on the right hand size, we now apply the operator
$U_G^m$ 
to the left.
In the first term, according to Eq.
(\ref{eq:getmsl})  	
this gives rise to the state
$\langle \eta |$.
In the second term
the state on the left hand size of 
$U_G^{\tau}$ 
is a fixed point of this operator and
thus remains unchanged.
Therefore 

\begin{equation}
\left| \langle m_1 | U_G^{\tau} |\phi\rangle \right|^2  
+ \left| \langle m_2| U_G^{\tau} |\phi\rangle \right|^2  
=
\left| \langle \eta |\phi\rangle \right|^2  
+ {\frac{1}{2}} \left| (\langle m_1 | - \langle m_2 |) \phi\rangle \right|^2.  
\label{eq:symasym2}
\end{equation}

\noindent
Note that the first term on the right hand size does not
depend on the choice of the two marked states, similarly
to what we obtained for a single marked state.
Therefore, the averaging over all the possible choices
of two marked states out of $N$ basis states is
not necessary.
However, the second term does depend on the
choise of
$| m_1 \rangle$
and 
$| m_2 \rangle$.
Thus the averaging is required, resulting in
   
\begin{equation}
P_{\rm s} =  
\left| \langle \eta | \phi\rangle \right|^2  
+ \frac{1}{N(N-1)} 
\sum_{m_1,m_2=0}^{N-1}
\left| (\langle m_1 | - \langle m_2) | \phi\rangle \right|^2,  
\label{eq:pmax5}
\end{equation}

\noindent
where the 
cases 
$m_1=m_2$
are excluded from the summation.
The first term on the right hand side is identical to the
expression obtained from $P_{\rm s}$ in the case of a single marked
state.
We will now show that $P_{\rm s}$ for two marked states is 
the same up to small corrections.
To this end we will put an upper bound on the value of the
second term in Eq.
(\ref{eq:pmax5}),
using the inequality

\begin{equation}
\left| (\langle m_1 | - \langle m_2 |) \phi\rangle \right|^2
\le
2 \left| \langle m_1  | \phi \rangle \right|^2 + 
2 \left| \langle m_2  | \phi \rangle \right|^2.
\end{equation}

\noindent
The double summation in the second term 
of Eq.~(\ref{eq:pmax5})
gives rise to sums of the form
$\sum_{i=0}^{N-1} |\langle i | \phi \rangle|^2 =1$.
The contribution of this term
is found to be just a correction of 
order $1/N$.  
We thus conclude that the probability of success 
$P_{\rm s}$
of the quantum search algorithm with the initial state
$| \phi \rangle$
and two marked states is identical to  
Eq.
(\ref{eq:pmaxetaphi}),
namely, it is the same probability as for a single marked state.

The result of this Section can be easily generalized to the case of $r$
marked states. In this case the second term on the right hand side
of Eq.
(\ref{eq:pmax5})
includes a sum over all the pairs of marked states.  
The conclusion that  
$P_{\rm s}$ 
is independent of the number of marked states
holds as long as $r \ll N$.

\subsection{Calculation of $P_{\rm s}$ for certain pure initial states}

Consider a state 
$| \phi \rangle$
with amplitudes $a_i$.
Since the amplitudes
$a_i = |a_i| e^{\theta_i}$, 
$i=0,\dots,N-1$
are complex numbers and satisfy the
normalization condition,
the value of
$|\bar a|$
increases as the amplitudes become more
alligned in the complex plane,
namely exhibit a narrow distribution of phases
$\theta_i$,
as well as a narrow distribution of the 
$|a_i|$'s.
As the distributions become broader the success probability
decreases.

For the state
$| \eta \rangle$,
in which all the amplitudes are
identical, with
$\bar a = 1/\sqrt{N}$,
the success probability is
$P_{\rm s} = 1$.
Consider a state 
$| \phi \rangle$
in which the amplitudes of
$k$ of the basis states 
are
$a_i = 1/\sqrt{k}$,
and for all the rest $N-k$
basis states
$a_i = 0$.
For such states the average amplitude
is 
$\bar a = \sqrt{k}/N$
and thus
$P_{\rm s} = k/N$.
Clearly, the success probability increases as the amplitude
is divided more evenly between the basis states.
There are several well known quantum states that can now be
examined.
One of them is the generalized GHZ state of $n$ qubits

\begin{equation}
| \phi \rangle_{\rm GHZ} = {\frac{1}{\sqrt{2}}}
(|00\dots0\rangle + |11\dots1\rangle).
\label{eq:ghz}
\end{equation}

\noindent
In this state only two of the amplitudes are non-zero.
Therefore the success probability
$P_{\rm s} = 2/N \rightarrow 0$,
as the number of qubits increases.
A similar situation is encountered for the
$W$ state of $n$ qubits given by

\begin{equation}
| \phi \rangle_{W} 
= {\frac{1}{\sqrt{n}}}
  (|10\dots0\rangle 
+ |010\dots0\rangle
+ \dots
+ |00\dots1\rangle).
\label{eq:W}
\end{equation}

\noindent
For this state
$P_{\rm s} = n/N$,
which also decays to zero as
$n$ is increased.

A large class of states for which
$P_{\rm s}$
vanishes
up to
$O(1/\sqrt{N})$
includes all the states
for which $\bar{a} = 0$.
This class includes, for
example, states in which for
each amplitude
$a_i$ of basis state $i$,
there is a state $j$, with
$a_j = - a_i$.
In general, a random sampling of states in the Hilbert
space of $n$ qubits tends to yield states with
very small $\bar a \simeq 1/N$.
This indicates that generic quantum states of $n$
qubits are highly inefficient as initial states for 
Grover's algorithm. 

\section{Discussion}

Recently it was shown that the success probability of
Grover's algorithm can be used in the construction of
an entanglement monotone that quantifies the entanglement
of pure quantum states of multiple qubits
\cite{Biham2002}.
To quantify the entanglement of the state
$| \phi \rangle$,
one is allowed to perform any set of local unitary operations
$U_1,U_2,\dots,U_n$
on the $n$ qubits before feeding the register
into the Grover apparatus.
These operations are designed to maximize the success probability
$P_{\rm s}$ of the algorithm.
For the case of a single marked state,
the resulting success probability will thus be
\cite{Biham2002}

\begin{equation}
P_{\max} = 
\max_{U_1,\ldots,U_n} 
\frac{1}{N} \sum_{m_1=0}^{N-1}
\left| \langle m_1 | U_G^{\tau} U_1 \otimes U_2 \otimes \cdots \otimes U_n
|\phi\rangle \right|^2  + O\left(\frac{1}{\sqrt{N}}\right),
\label{eq:pmax8}
\end{equation}

\noindent
where the summation takes care of the averaging over
all possible choices of the marked state
$| m_1 \rangle$
and the maximization is over all possible sets
of local unitary operators
$U_i$, $i=1,\dots,n$.
It was shown that 
$P_{\rm max}$
can be reduced to the form

\begin{equation}
P_{\max} = 
\max_{|s_1, \ldots, s_n\rangle} 
\left| \langle s_1,\ldots, s_n|\phi \rangle \right|^2 +
O\left(\frac{1}{\sqrt{N}}\right),
\label{eq:pmax9}
\end{equation}

\noindent
where the maximization now runs over all product states,
$|s_1,\ldots,s_n\rangle = |s_1\rangle\otimes\cdots|s_n\rangle$, 
of the $n$ qubits.
The maximal success probability is thus determined by the
maximal overlap between 
$| \phi \rangle$
and any product state of 
$n$ qubits.
The Groverian entanglement measure was defined
as

\begin{equation}
G(\phi) = \sqrt{1-P_{\rm max}}.
\end{equation}

\noindent
It was shown that 
$G(\phi)$
is an entanglement monotone
\cite{Biham2002,Vedral1997}.
This means that
for product states
$G(\phi) = 0$,
while for any state
$|\phi\rangle$
of $n$ qubits
it is invariant under any local unitary operations
on single qubits.
Moreover, 
$G(\phi)$
cannot be increased by 
{\it any}
local operations on the $n$
qubits (where classical communication is allowed between the
parties that perform these operations).

The result presented above, that the success probability
$P_{\rm s}$ does not depend on the number of marked states,
provides a generalization of the Groverian entanglement. 
It removes the restriction that 
$G(\phi)$
will be defined by a search with a single marked state.
The use of the Groverian entanglement measure
to characterize entangled quantum states of
multiple qubits may provide useful insight about the nature
of these states and their role in quantum algorithms. 
To this end
one needs to develop efficient computational schemes
for the maximization 
over all possible sets
of local unitary operations
$U_i$, $i=1,\dots,n$
[Eq. (\ref{eq:pmax8})],
or, alternatively,
over all sets of product states
[Eq. (\ref{eq:pmax9})].

\section{Summary}

We have analyzed the dynamics of
Grover's quantum search algorithm
for the case in which
the initial state 
is an arbitrary pure quantum state 
$| \phi \rangle$ 
of $n$ qubits. 
We have 
shown that the optimal time to perform the measurement
that concludes the operation of the algorithm
is independent of the initial state
$| \phi \rangle$.
It is identical to the optimal measurement time of the original algorithm,
with the same number of marked states, 
in which
$| \phi \rangle = | 0 \rangle$.
An expression for 
the probability of success 
$P_{\rm s}$ 
in terms of the amplitudes of the state
$| \phi \rangle$
is obtained
and is shown to be independent of the number of marked states, $r$.   
The fixed points and cycles of the Grover
operator $U_G$ are identified.
The relevance of
the results in the context of using the success probability as an
entanglement measure is discussed.
In particular, the Groverian entanglement measure,
previously limited to a single marked state,
is generalized to the case of several marked states.
It is shown that as long as $r \ll N$,
$G(\phi)$ is independent of $r$.

\acknowledgments

O.B would like to thank M.A. Nielsen, T.J. Osborne and 
J. Dodd for discussions during his sabbatical at the University
of Queensland in Summer 2001, that stimulated this work.
We acknowledge financial support from the
EU, under Grant No.~IST-1999-11234.

\end{document}